\documentclass[conference]{IEEEtran}
\ifCLASSINFOpdf
   \usepackage[pdftex]{graphicx}
   \DeclareGraphicsExtensions{.pdf,.jpeg,.png}
\else
\fi
\ifCLASSOPTIONcompsoc
  \usepackage[caption=false,font=normalsize,labelfont=sf,textfont=sf]{subfig}
\else
  \usepackage[caption=false,font=footnotesize]{subfig}
\fi
\begin{document}
%
\title{The WaveDAQ integrated Trigger and Data Acquisition System for the MEG II experiment}



%
\author{\IEEEauthorblockN{Marco Francesconi\IEEEauthorrefmark{1}\IEEEauthorrefmark{2},
Alessandro Massimo Baldini\IEEEauthorrefmark{1}\IEEEauthorrefmark{2},
Fabrizio Cei\IEEEauthorrefmark{1}\IEEEauthorrefmark{2},
Marco Chiappini\IEEEauthorrefmark{5}\IEEEauthorrefmark{2},
Luca Galli\IEEEauthorrefmark{2},\\
Marco Grassi\IEEEauthorrefmark{2},
Ueli Hartmann\IEEEauthorrefmark{3},
Manuel Meucci\IEEEauthorrefmark{4},
Fabio Morsani\IEEEauthorrefmark{2},
Donato Nicol\`o\IEEEauthorrefmark{1}\IEEEauthorrefmark{2},
Angela Papa\IEEEauthorrefmark{1}\IEEEauthorrefmark{2}\IEEEauthorrefmark{3},\\
Stefan Ritt\IEEEauthorrefmark{3},
Elmar Schmid\IEEEauthorrefmark{3} and
Giovanni Signorelli\IEEEauthorrefmark{2} }
\IEEEauthorblockA{\IEEEauthorrefmark{1}Dipartimento di Fisica dell'Universit\`a degli Studi di Pisa,
Largo B. Pontecorvo 3, 56127 Pisa, Italy}
\IEEEauthorblockA{\IEEEauthorrefmark{2}INFN Sezione di Pisa
Largo B. Pontecorvo 3, 56127 Pisa, Italy}
\IEEEauthorblockA{\IEEEauthorrefmark{3}Paul Scherrer Institut PSI
5232 Villigen, Switzerland}
\IEEEauthorblockA{\IEEEauthorrefmark{4}INFN Sezione di Roma
Piazzale A. Moro, 00185 Rome, Italy}
\IEEEauthorblockA{\IEEEauthorrefmark{5}Dipartimento di Scienze Fisiche,
della Terra e dell'Ambiente dell'Universit\`a, via Roma 56, 53100, Siena, Italy}
}


\maketitle

\begin{abstract}
The WaveDAQ is a newly-designed digitization Trigger and Data AcQuisition system (TDAQ) allowing Multi-gigasample waveform recording on a large amount of channels (up to 16384) by using the DRS4 analog switched capacitor array as downconverting ASIC.
A high bandwidth, programmable input stage has been coupled with a bias generator to allow SiPM operation without need of any other external apparatus. The trigger generation is tightly coupled within the system to limit the required depth of the analog memory, allowing faster digitization speeds.
This system has been designed for the MEG experiment upgrade but also proved to be highly scalable and already found other applications.
\end{abstract}


%
\IEEEpeerreviewmaketitle

\section{Introduction}
The MEG experiment \cite{meg_final} pioneered the use of multi-gigasample digitizer to achieve the stringent timing and charge resolutions needed to disentangle the high pileup environment the detectors face.
In the framework of the global upgrade plan \cite{upgrade}, that aims to a factor 10 higher sensitivity on $\mu^+ \rightarrow e^+ \gamma$ (current limit was set by MEG to $BR < 4.2\cdot10^{-13}$ at 90\% CL), a global redesign of the trigger and data acquisition systems was needed. In particular the number of channels increased by a factor of 4 by switching from PMTs to SiPMs while the electronics had to fit in the same rack space. Built on the experience gained with the former Trigger \cite{meg_trigger_hw} and DAQ \cite{DRS4} systems, the new WaveDAQ can exploit the features of DRS4 analog switched capacitor array to an unprecedented level, in particular the readout time will be decreased by moving from the VME standard to gigabit ethernet interfaces and and by merging the two former systems in a single, tightly integrated system to ease the interplay.

\section{System description}
\begin{figure*}[!t]
\centering
\includegraphics[width=5.1in]{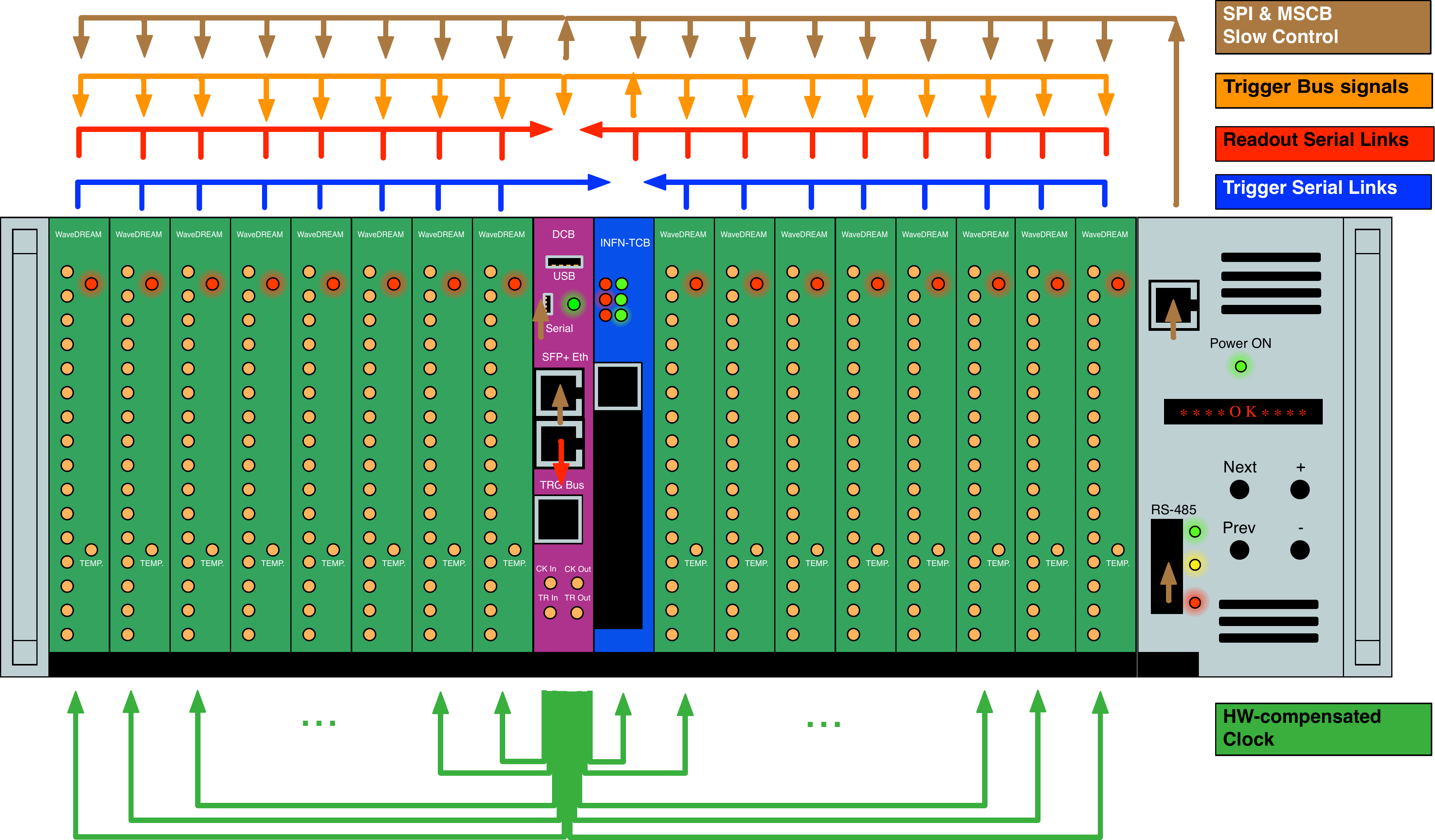}
\caption{Sketch of a single WaveDAQ crate: green boards are WaveDREAM frontend boards, the magenta board is the Data Concentrator Board and the blue board is the Trigger Concentrator Board. Arrows show connections in the backplane: red arrows for data transmission to backend machines, blue arrows for trigger serial links, orange arrows to distribute back the trigger signal and green arrows for hardware compensated clock distribution. Brown arrows show low level access for slow control and configuration.}
\label{Crate}
\end{figure*}

While designing the new system no available standard delivered the required compactness, connection topology and clock distribution requirements. Therefore, a system with a fully custom backplane had to be designed. This system features a fully custom backplane with dual-star serial link capability and a skew-compensated low jitter clock distribution.
In a standard 3U rack format, 16 slave boards are connected to the two central slots for trigger and data processing as shown in figure \ref{Crate} where we also highlighted the various connections between the boards fully described in the next sections.

The crate housekeeping is accomplished by a microcontroller board integrated into the 360W crate power supply, rightmost in picture \ref{Crate}, that handles the shared power lines and the temperature-controlled fans while providing a low level interface to the boards in the crate for slow control and configuration though a dedicated Ethernet network connection.

\subsection{WaveDREAM Frontend Board}
\begin{figure}[!t]
\centering
\includegraphics[width=2.5in]{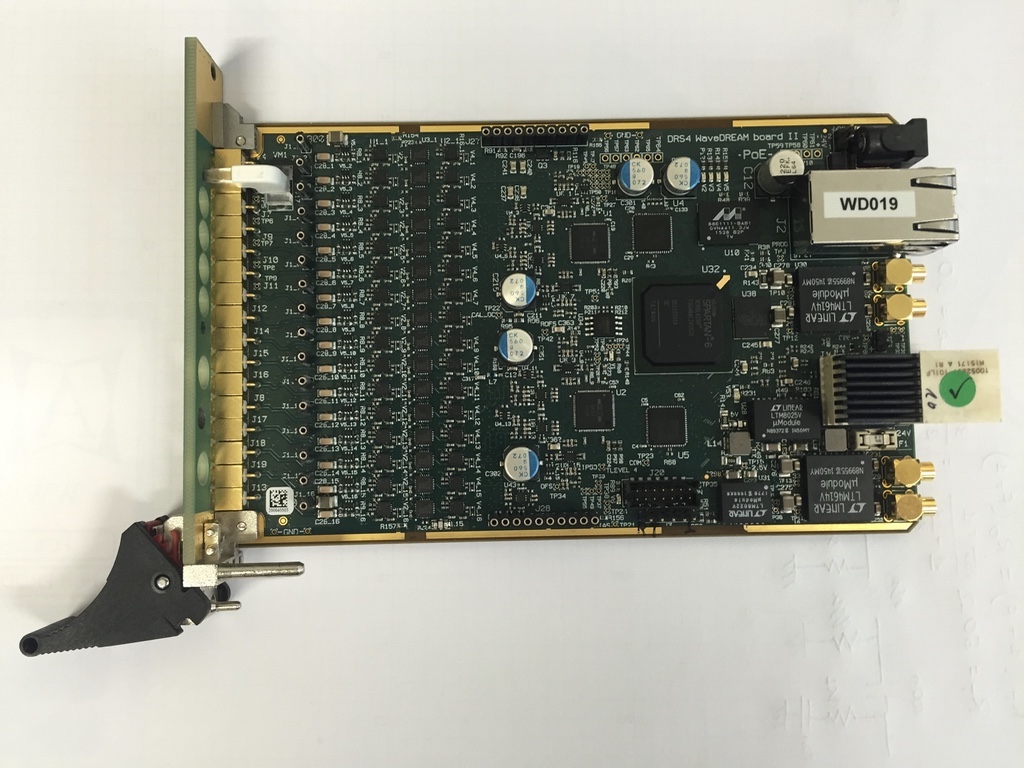}
\caption{WaveDREAM board housing DRS4 digitizers, ADCs and TDCs for trigger. The ethernet plug on the back may not be populated when used inside a crate. The piggy back HV module is not shown in picture.}
\label{WDB}
\end{figure}

Usually slave positions in a crate are occupied by the Waveform Drs4 REadout Module (WaveDREAM), pictured in figure \ref{WDB}. That board contains two DRS4 chips, capable of 0.5-5 GSPS sampling speeds, together with 900MHz bandwidth variable gain amplifiers to allow direct connection of detector signals.
The same 80MSPS ADC used to digitize analog amplitudes stored in the DRS4 will provide also an independent sampling while the analog digitizer is running, to be used inside the trigger logic in combination with fast analog comparators.
The information from the analog comparators is also fed to a time to digital conversion logic implemented in the readout FPGA by means of high speed shift registers to result in low resource usage while achieving 450ps time resolution, limited by the fastest available clock in the FPGA.
A voltage bias generator may provide channel-by-channel biasing for single SiPM or for multi-SiPM arrays without requiring any external power supply.
When inside the crate up to 256 channels on 16 boards can be digitized simultaneously.

The WaveDREAM can be used outside a crate using the onboard Gbit Ethernet interface. In this configuration a single board provides a benchtop DAQ platform with 16 acquisition channels.

\subsection{Data Concentrator Board and Trigger Concentrator Board}
\begin{figure*}[!t]
\centering
\subfloat[DCB]{\includegraphics[width=2.5in]{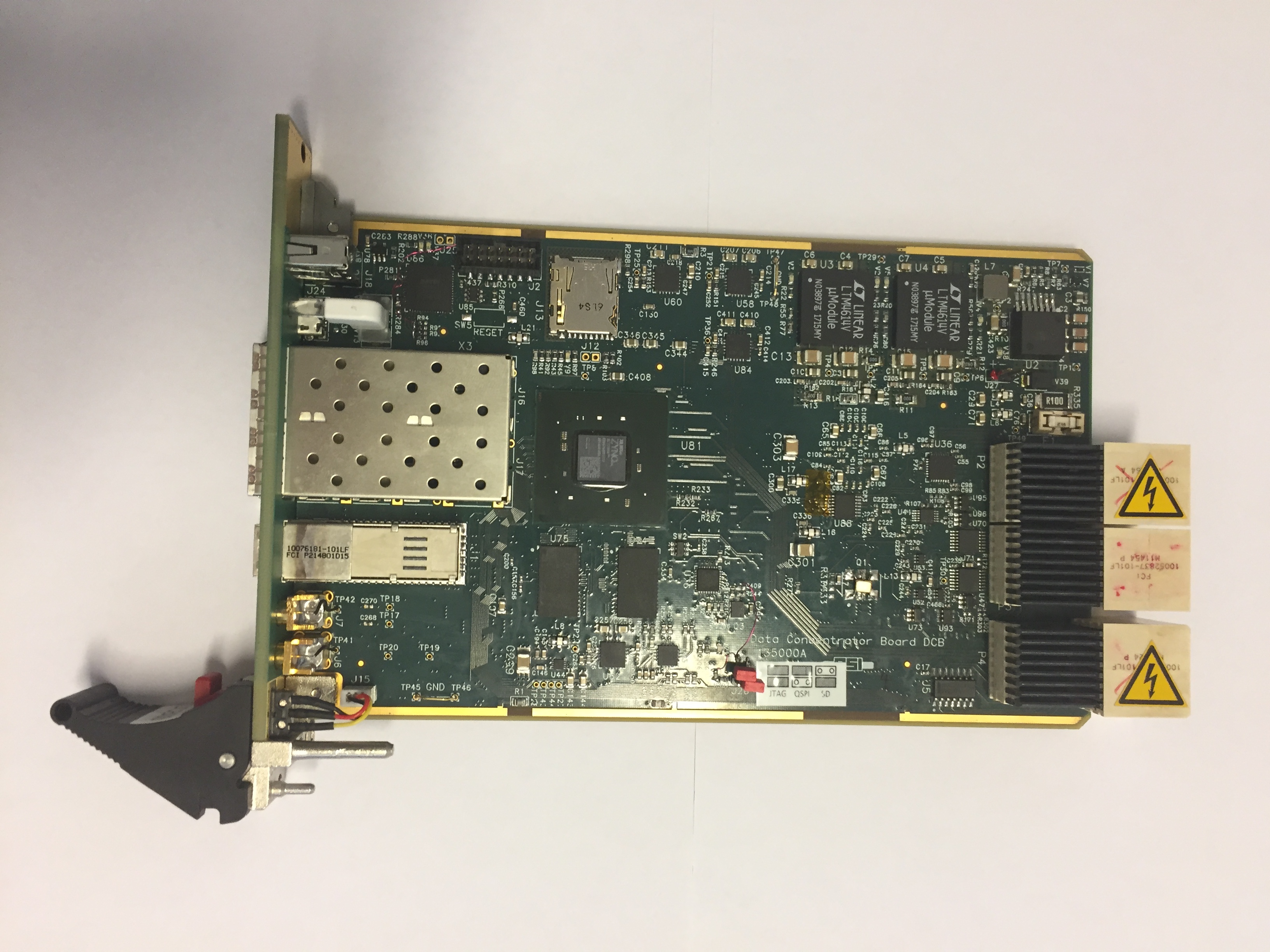}%
\label{DCB}}
\hfil
\subfloat[TCB]{\includegraphics[width=2.5in]{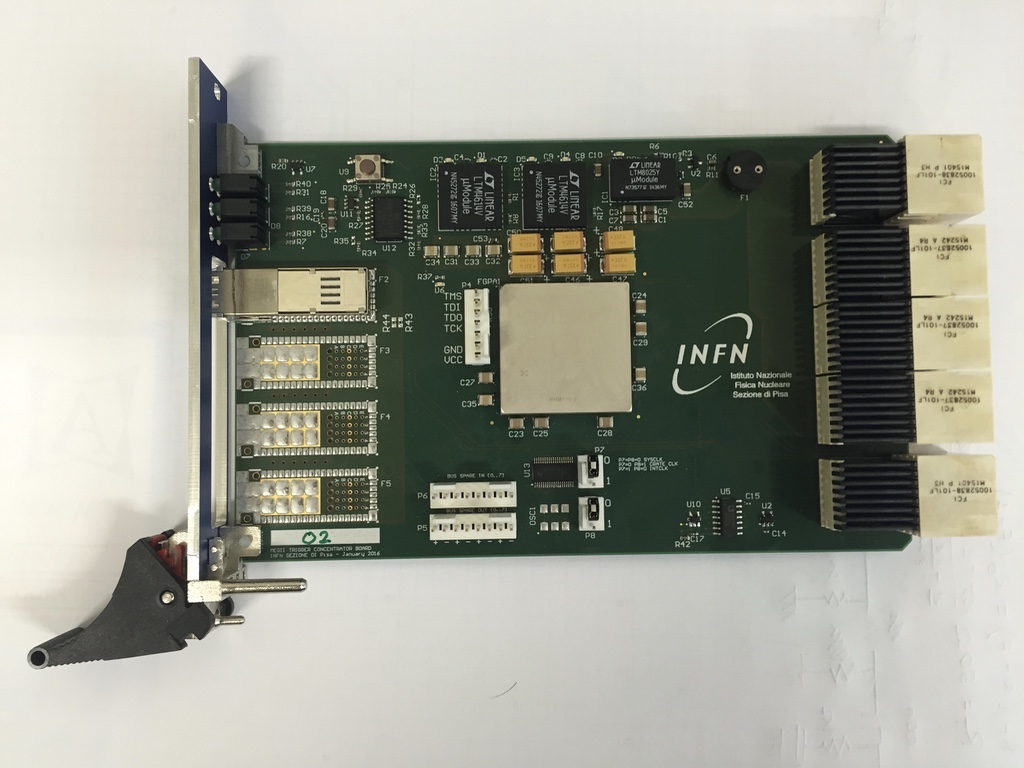}%
\label{TCB}}
\caption{Data Concentration Board \ref{DCB} and Trigger Concentration Board \ref{TCB}.}
\label{others}
\end{figure*}

The two central slots in a crate are reserved for custom designed boards that take care of crate-level functions mandatory for multi-board DAQ operation:
\begin{itemize}
\item The Data Concentrator Board (DCB), figure \ref{DCB}, generates and distributes the main reference clock while handling the data received by WaveDREAMs, combining them into a single Gigabit Ethernet interface. The onboard Zynq SystemOnChip acts as master of the crate, handling the interface to the other boards for configuration.

\item The Trigger Concentrator Board (TCB), figure \ref{TCB}, receives all trigger informations from the WaveDREAMs using low-latency serial links to generate a shared trigger signal when amplitude and time based algorithms detect an event of interest. All online selection strategies for MEG II will be based on the ones used in MEG \cite{meg_trigger_fw} with the necessary modifications and improvements by the different detector segmentations.
\end{itemize}

\subsection{Use in experiments}
In experiments which require more than 256 channels, a shared trigger generation and distribution is necessary.
trigger generation, a dedicated so-called trigger concentrator crate crate filled with TCBs in slave and master positions can collect data from TCBs sitting in up to 64 crates by using high speed serial cables on the front panel. The TCB in the master position of the trigger concentration crate can receive trigger informations from all boards in the system and eventually generate the trigger signal.

Another crate, using fanout cards, will account for the clock generation and the trigger signal distribution.
Using such a design 16384 channels can fit in ten 21U racks, resulting in more than 4 times the channel density achieved in MEG.

\section{Prototype operation}
\begin{figure}[!t]
\centering
\includegraphics[width=2.5in]{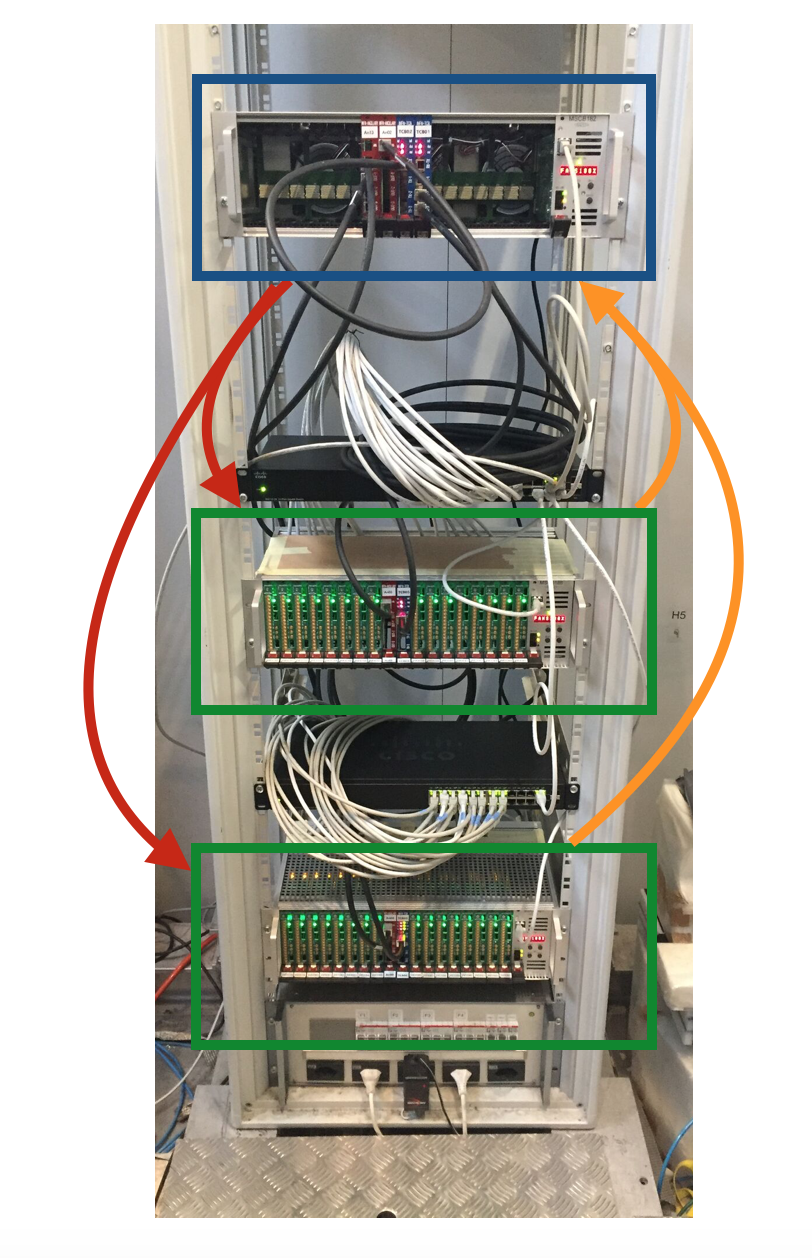}
\caption{Photo of a prototype crate setup involving two WaveDREAM Crates (green) and a trigger computation and distribution crate (blue). Orange arrows show the trigger data flow on cables toward the trigger crate while red arrows are for trigger signal distribution. }
\label{System}
\end{figure}
Several tests with prototype systems were performed during the last two years, ranging from single board tests, for signal checks or small detector tests, to 6-crate multi-detector systems used during beam tests of the upgraded detectors.
One of those systems is shown in figure \ref{System}. Even if smaller than the final MEG II system, lots of features such as board synchronization, busy handling and trigger communications have been studied and optimized in this setup before the final system production.

The two crate system of picture \ref{System} showed very good timing characteristics when tested with the new MEG II Timing Counter detector\cite{new_tc}, made by scintillating tiles readout by SIPMs: the electronics contribution have been measured to be of the order of 10 ps and have proved to be stable over a few month period. During the beam test period the Timing Counter detector has achieved its goal of about 35ps time resolution on positron tracks without being limited by digitizer performances (for details see \cite{upgrade}).

\begin{figure}[!t]
\centering
\includegraphics[width=2.5in]{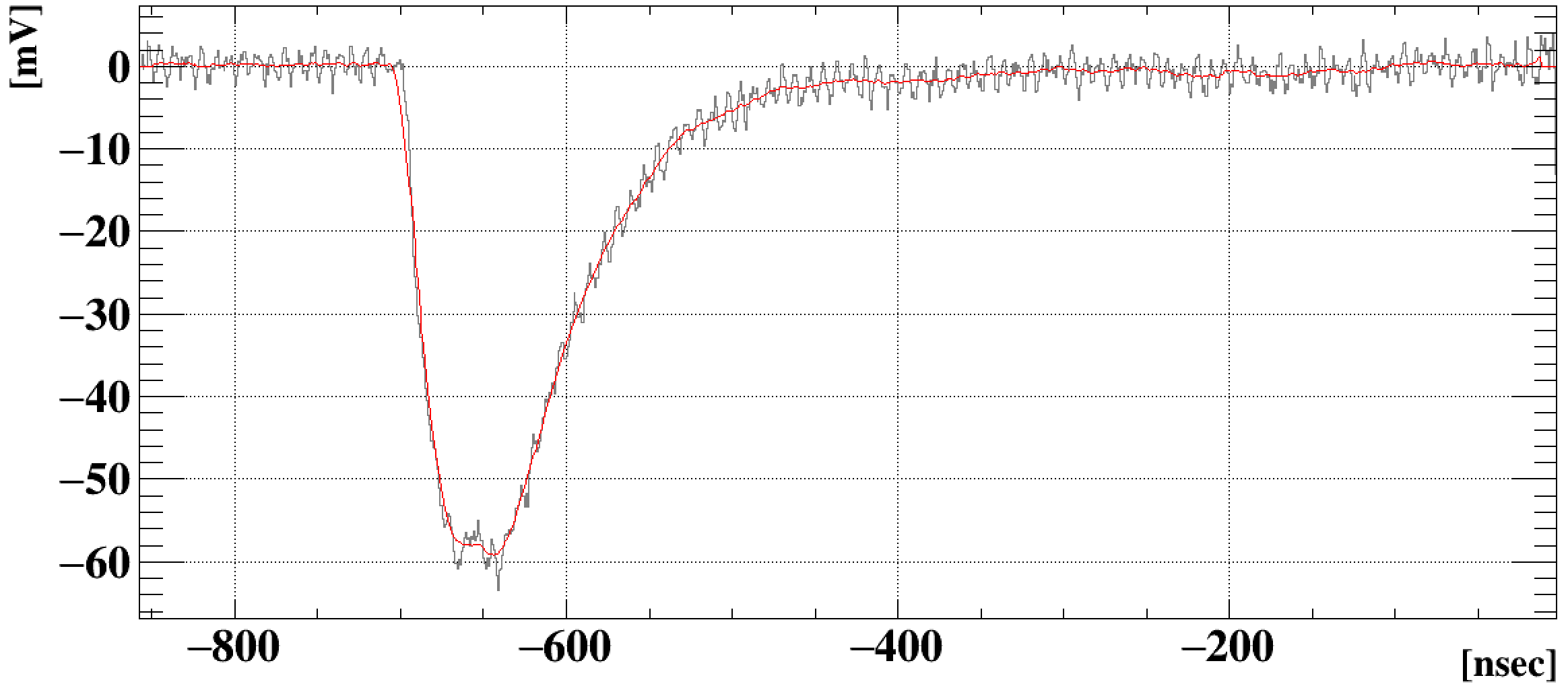}
\caption{Example Waveform recorded by DRS4 with the MEG II Liquid Xenon Detector. The trigger is at zero in the horizontal axis, so the overall trigger latency is $\sim700ns$}
\label{Waveform}
\end{figure}

During the same test beam four crates have been devoted to readout a subset of the Liquid Xenon detectors \cite{new_xec}, so that the full trigger reconstruction, involving pedestal-subtraction, weighted sum and discrimination with veto, has been tested and proved to generate a trigger within 700ns, as visible from the recorded DRS4 waveform shown in figure \ref{Waveform}. Using such logic we succeeded in selecting beam-induced photon events, cosmics, and other photon calibration sources.
The latency will be improved by $\sim200ns$ in the next WaveDREAM board revision by using an ADC with a shorter pipeline.

Outside the MEG experiment, a single crate system has been chosen to readout the dE/TOF Detector of the FOOT Experiment \cite{FOOT}, few prototypes have been tested with single-board DAQ, achieving the required timing and charge performance.

\section{Conclusion}
We showed a new digitizing Trigger and Data Acquisition system being assembled for the MEG II experiment. The system allows the use of several DRS4 digitizers on a large number of channels and is easy to scale from a 16 channel single board setup to $\sim16000$ channels. Several configurations involving smaller size systems have been tested and all underlying functionalities have been proved to be operational.




\begin{thebibliography}{1}

\bibitem{meg_final}
J. Adam et al., The MEG detector for $\mu^+ \rightarrow e^+\gamma$ decay search. Eur. Phys. J. C 73, 2365 (2013). https://doi.org/10.1140/epjc/ s10052-013-2365-2. arXiv:1303.2348

\bibitem{upgrade}
A. Baldini et al., The design of the MEG II experiment, 2013, Eur. Phys. J. C (2018) 78:380 https://doi.org/10.1140/epjc/s10052-018-5845-6

\bibitem{meg_trigger_hw}
L. Galli et al., An FPGA-based trigger system for the search of μ → eγ decay in the MEG experiment. J. Instrum. 8, P01008 (2013). doi:10.1088/1748-0221/8/01/P01008

\bibitem{DRS4}
S. Ritt, R. Dinapoli, U. Hartmann, Application of the DRS chip for fast waveform digitizing. Nucl. Instrm. Methods. A 623(1), 486488 (2010). doi:10.1016/j.nima.2010.03.045

\bibitem{meg_trigger_fw}
L. Galli et al., Operation and performance of the trigger system of the MEG experiment. J. Instrum. 9, P04022 (2014). doi:10.1088/ 1748- 0221/9/04/P04022

\bibitem{new_tc}
Y. Uchiyama et al., 30-ps time resolution with segmented scintillation counter for MEG II. Nucl. Instrum. Methods A 845, 507–510 (2017).  https://doi.org/10.1016/j.nima.2016.06.072

\bibitem{new_xec}
W. Ootani et al., Development of deep-UV sensitive MPPC for liquid xenon scintillation detector. Nucl. Instrum. Methods A A787, 220–223 (2015). https://doi.org/10.1016/j.nima.2014.12.007

\bibitem{FOOT}
V. Patera, et al. ”The FOOT (Fragmentation Of Target) Experiment.” PoS (2017): 128. doi:10.22323/1.281.0128


\end{thebibliography}
\end{document}